\documentstyle[titlepage]{article}
\newcommand{\R}{{\rm I\kern-2pt R}}

\begin{document}
\begin{center}
{\bf Note on the Controllability of Molecular Systems\\}
Viswanath Ramakrishna\\
Department of Mathematical Sciences and Center for Signals, Systems
and Telecommunications\\
University of Texas at Dallas\\   
Richardson, TX 75083 USA\\
email: vish@utdallas.edu\\
Supported in part by NSF - DMS-0072415 
\end{center}

\noindent{ \bf ABSTRACT:}
The purpose of this brief note is to show that the set of states reachable
from a given initial state for a finite dimensional quantum system is
equal to the orbit under the Lie group generated by the 
Lie algebra generated by the internal and external Hamiltonians,
{\it even if this Lie group is not compact.} This result has been
explicitly mentioned in several talks by the author. This note is not
being submitted for publication in any journal.

\section{A Slight Improvement}
 Consider a finite- dimensional quantum system:
\begin{equation}
\dot{c} = Ac + Bc\epsilon (t), c(0) = c_{0}
\end{equation}

Here $c$ evolves on the complex sphere in some $C^{N}$, and $A$ and
$B$ are two skew-Hermitian matrices. 
It is well known that the set of states reachable from $c_{0}$, via the
application of admissible controls, equals the orbit through
$c_{0}$ of the group, $G$ corresponding to the Lie algebra 
generated by
the matrices $A$ and $B$, {\it provided this Lie group is compact.}
See \cite{hector,pra,sonia} for instance. {\it This result is valid even if
$G$ is not compact.} This result has been mentioned explicitly by
the author in several talks and grant proposals,
though the result has been recorded only
cryptically in \cite{gnvn99detroit,gnvn99tlaxcala} (primarily because
it  offers no significant practical improvements). Since such a nominal
improvement may, nevertheless, be of interest, a discussion of it is
provided below.

To that end, notice that $A$ may be assumed to be diagonal,
$A = {\mbox diag} \ (i\lambda_{1}, i\lambda_{2}, \ldots , i\lambda_{N})$.
Writing $c_{k} = a_{k} + ib_{k}, k=1, \ldots , N$ and viewing the given
quantum system as evolving on $R^{2N}$, it follows that the drift portion
of the system, i.e., the vector field $Ac$, is a classical Hamiltonian system
with Hamiltonian, $H = \sum_{k=1}^{N} \lambda_{k} (a_{k}^{2} + b_{k}^{2})$
(under the standard symplectic structure of $R^{2N}$).

Now the quantum system evolves on the sphere $S = \{ \sum_{i=1}^{N}
\mid c_{i}\mid^{2} \  = 1 \}$. $S$ is of course also the set
$\{ \sum_{i=1}^{N} a_{i}^{2} + b_{i}^{2} \ = 1\}$. Thus $S$ is a bounded
set and further the orbit of $G$ through the initial condition in $S$
is also bounded, even if it is not compact. Thus, the quantum system
can be viewed as a nonlinear control system on a bounded manifold,
whose drift is a classical Hamiltonian system. Since this implies that
the drift is weakly positively Poisson stable, the desired conclusion
follows from
\cite{yang}(see also \cite{daya}). Indeed, these articles show that, under the 
weak positively Poisson stability of the drift, the reachable set 
from the initial condition is exactly the leaf of the foliation associated
to the drift and control vector fields. But this leaf is nothing but the
orbit of $G$ through the initial condition.   
It also implies that the reachable set of the associated unitary generator
system from the identity matrix is $G$ (cf., \cite{thug}),
since if this is not the case, the
reachable set of the quantum system could not be the orbit of $G$ through
the initial condition.

In the opinion of the author, this improvement is of mathematical 
interest only. The result is of relevance in the situation that the
quantum system is not completely controllable, i.e., when the orbit of
$G$ is not all of $S$. However, for the result to have practical value
it is desirable to describe the orbit explicitly. One way to do this
is to use invariant theory and describe the orbit as a level set.
However, invariant theory is tractable
only for compact and reductive groups. The author is not aware of any
practical situation in which $G$ is not compact but is reductive.

\end{document}